\documentstyle[12pt,epsf,axodraw]{article}
\makeatletter
\def\appendix{\par\clearpage
  \setcounter{section}{0}
  \setcounter{subsection}{0}
  \@addtoreset{equation}{section}
  \def\@sectname{Appendix~}
  \def\theequation{\thesection.\arabic{equation}}
  \def\thesection{\Alph{section}}}
\makeatother

\begin{document}

\begin{titlepage}
\hskip 11cm \vbox{\hbox{BUDKERINP/01-32}\hbox{DFCAL-TH 01/3}
\hbox{June 2001}}
\vskip 0.3cm
\centerline{\bf CALCULATION OF REGGEON VERTICES IN QCD$^{~\ast}$}
\vskip 1.0cm
\centerline{  V.S. Fadin$^{a~\dagger}$ and R. Fiore$^{b~\ddagger}$ }
\centerline{\sl $^{a}$ Budker Institute for Nuclear Physics,}
\centerline{\sl Novosibirsk State University, 630090 Novosibirsk, Russia}
\vskip 0,5cm
\centerline{\sl $^{b}$ Dipartimento di Fisica, Universit\`a della Calabria,}
\centerline{\sl Istituto Nazionale di Fisica Nucleare, Gruppo collegato di Cosenza,}
\centerline{\sl I-87036 Arcavacata di Rende, Cosenza, Italy}
\vskip 1cm

\begin{abstract}
The method of calculation of effective vertices of interaction of
the  Reggeized gluon and quark with particles in QCD in the
next-to-leading order is developed. The method is demonstrated in
the case of already known vertices of both gluon-gluon and
quark-quark transitions in the scattering of gluons and quarks on
the Reggeized gluon. It is used for the calculation of the
gluon-quark transition in the scattering on the Reggeized quark.
\end{abstract}
\vfill

\hrule
\vskip.3cm

\noindent
$^{\ast}${\it Work supported in part by the Ministero italiano
dell'Universit\`a e della Ricerca Scientifica e Tecnologica and in part by
the Russian Fund of Basic Researches.}
\vfill
$ \begin{array}{ll}
^{\dagger}\mbox{{\it email address:}} &
 \mbox{FADIN~@INP.NSK.SU}\\
\end{array}
$

$ \begin{array}{ll}
^{\ddagger}\mbox{{\it email address:}} &
  \mbox{FIORE~@FIS.UNICAL.IT}
\end{array}
$
\vfill
\vskip .1cm
\vfill
\end{titlepage}
\eject
\textheight 210mm \topmargin 2mm \baselineskip=24pt

\newpage

\section{Introduction}
Investigation of a possibility of Reggeization of elementary
particles in field theories was started in Ref.~\cite{GGLMZ},
where it was shown that the electron in Quantum Electrodynamics
(QED) does Reggeizes in perturbation theory. Soon after it was
proved~\cite{M} that, contrary to the electron, the photon in QED
remains elementary. The problem of Reggeization of elementary
particles in non-Abelian gauge theories (NAGT's) naturally
appeared with the development of these theories. In
Ref.~\cite{GST} it was found that the criteria of Reggeization
formulated by Mandelstam ~\cite{M} are fulfilled in NAGT's for
all particles. The notion ``Reggeization'' was used in
Refs.~\cite{GGLMZ,GST} for the disappearance, because of
radiative corrections, of the non-analytic terms in the complex
angular momentum plane, related to elementary particle exchanges
in the Born approximation. In Refs.~\cite{L76,BFKL} it was shown by direct
calculations in the leading logarithmic approximation (LLA) that,
for the gauge bosons of NAGT's, this term can be understood in a
much stronger sense. That means not only the existence of the
Reggeon with the quantum numbers of the gauge boson, negative
signature and trajectory $j(t) = 1+\omega (t)$ passing through
$1$ \ \ at $\ t = m_{V }^{2}$, \ $m_{V }$\ \ \ being a gauge boson
mass, but also that in each order of perturbation theory at large 
c.m.s. energies $\sqrt{s}$ \ this Reggeon gives the leading contribution
to the scattering amplitudes with the quantum numbers of the gauge
boson and the negative signature in the $t$-channel. Below we use the
term ``Reggeization'' in such strong sense. In Ref.~\cite{FS}
it was demonstrated, also by direct calculations in the LLA, the
Reggeization of fermions in NAGT's. Therefore, in Quantun
Chrodynamics (QCD), which is a particular case of NAGT, all
elementary particles, i.e. quarks and gluons, do Reggeize.

The Reggeization of elementary particles plays a very important
role for the description of high energy processes. The gluon
Reggeization is the basis of the famous BFKL
equation~\cite{BFKL}. The Pomeron, which determines the high
energy behaviour of cross sections, in QCD is a compound state of
two Reggeized gluons. The Odderon, responsible for the difference
of particle and antiparticle cross sections, can be constructed
as a compound state of three Reggeized gluons~\cite{L93}. One
could also construct colorless objects from Reggeized quarks and
antiquarks which should be relevant to phenomenological Reggeon
trajectories ssuccessfully used for the description of processes
with exchange of  quantum numbers.

For phenomenological applications the LLA is not satisfactory,
since neither the scale of energy, nor the scale of momentum
transfers in the argument of the coupling constant are fixed in
this approximation. The calculation of radiative corrections to
the kernel of the BFKL equation has taken many years of a hard
work~[8-13]. Three years ago the kernel of the BFKL equation was
obtained at the next-to-leading order (NLO)~\cite{FL98} for the
case of the forward scattering, i.e. the momentum transfer $t =
0$ and the vacuum quantum numbers in the $t$-channel. Although in
the $\overline{MS}$ renormalization scheme with a ``reasonable''
scale setting for the running QCD coupling radiative
corrections appear to be very large, use of non-abelian
renormalization schemes and the BLM procedure for the scale
setting opens a way for applications of the NLO BFKL 
in the high energy phenomenology~\cite{BFKLP}. Good possibilities
for applications to deep inelastic processes are given by the
renormalization group improvement of the BFKL equation~\cite{CCP}.

Due to the Reggeization of quarks and gluons, an important role
in high energy QCD belongs to the vertices of Reggeon-particle
interactions. In particular, these vertices are necessary for the
determination of the BFKL kernel. For their calculation powerful 
methods based on analyticity and unitarity were developed starting 
from the LLA~\cite{L76,BFKL}. They were intensively used for the 
calculation of the Reggeon-particle interaction vertices in the 
NLO. But these methods are not adjusted for the direct calculation 
of the vertices, which  are obtained from the comparison 
of appropriate scattering amplitudes with their Reggeized form, so 
that to obtain a vertex one must calculate a whole amplitude. This 
seems to be too complicated, and a method of calculation of Reggeon
vertices themselves is desirable. In principle, it could be  based on the 
effective action for the interactions of the Reggeized quarks and gluons 
with the usual QCD partons~\cite{L95,LV01}. However, a straightforward 
application of the effective action leads to vertices depending on the 
auxiliary parameter $\eta$ in the rapidity space~\cite{L95}, serving by 
a cut-off in the relative longitudinal momenta of the produced
particles. In order to find the correspondence between these vertices 
and the conventional ones we need again to know the whole amplitudes. 
Therefore we adopt the approach based on the properties of the 
integrals corresponding to the Feynman diagrams 
with two particles in the $t$-channel, noticed in Ref.~\cite{FM98}. It was
already used in Ref.~\cite{FM98} for the definition of the vertex for the
quark-antiquark production in the interaction of the virtual photon with
the Reggeized gluon.

In this paper we develop the method of the direct calculation of the
Reggeon-particle vertices. In the next Section we explain the essence of
the method both in the case of the Reggeized gluon (Subsection 2.1) and the
Reggeized quark (Subsection 2.2). In Section 3 we show applications
of the method. In Subsection 3.1 the quark-quark-Reggeon (QQR) vertex is
considered. Subsection 3.2 is devoted to the more complicated case of the
gluon-gluon-Reggeon (GGR) vertex. In Subsection 3.3 we calculate the 
vertices for the case in which the Reggeon is the Reggeized quark. Finally,
Section 4 contains a discussion of the method  and the obtained results.

\section{Method of calculation}
\subsection{Reggeized gluon vertices}
Let us consider the amplitude of  the process $A + B
\rightarrow A^{'} + B^{'}$ at large  c.m.s. energy squared $s=(p_{A}+p_{B})^{2}$ and
fixed momentum transfer squared $t=(p_{A^{\prime
}}-p_{A})^{2}=(p_{B}-p_{B^{\prime }})^{2}$ supposing that the 
one-gluon state is possible in the $t$-channel. Then the
projection of this amplitude on the color octet state in the
$t$-channel taken with the negative signature (i.e. antisymmetrized
with respect to $s \leftrightarrow u \approx -s)$ has the form
\begin{equation}
A_{8^-}=\Gamma _{A^{\prime }A}^{i}\left[ \left(
\frac{-s}{-t}\right) ^{j(t)}-\left( \frac{s}{-t}\right)
^{j(t)}\right] \Gamma _{B^{\prime }B}^{i}~, 
\label{z1}
\end{equation}
where $\Gamma _{A^{\prime }A}^{i}$ are the Reggeon vertices for the 
$A \rightarrow A^{'}$ transitions  and $j(t) = 1+\omega(t)$ is the gluon 
trajectory. In the leading order
\begin{equation}
\omega (t)=\omega ^{(1)}(t)=\frac{g^{2}t}{(2\pi
)^{D-1}}\frac{N}{2}\int
 \frac{d^{D-2}k_{\perp}}{k_{\perp}^{2}(q-k)_{\perp}^{2}}=
-\frac{g^{2}N\Gamma (1-\epsilon )(\vec q^{~2})^{\epsilon }}{(4\pi
)^{2+\epsilon }}\frac{\Gamma (\epsilon )}{\Gamma (2\epsilon )}~.
\label{z2}
\end{equation}
Here $t=q^2\approx q_{\perp}^{2}=-\vec q^{~2}$, the subscript
$\perp$ denotes components transverse to the plane of initial
momenta (we use also the vector sign for these components), $N$ is
the number of colours ($N$=3 in QCD) and $D=4+2\epsilon$ is the
space-time dimension taken different from 4 for regularization. The 
Reggeon vertices describing the quark-quark and gluon-gluon transitions 
in the leading order and in the helicity basis are quite simple:
\begin{equation}
\Gamma _{A^{\prime }A}^{(0)i}=g\delta _{
\lambda_{A^{\prime}}\lambda_{A}   } \langle A^{\prime }|T^{i}|A
\rangle~, 
\label{z3}
\end{equation}
where $\langle A^{\prime }|T^{i}|A \rangle$ stands for the  matrix
element of the colour group generator in the corresponding
representation. It is necessary to note that the form (\ref{z1}) has a
general nature and is valid not only for such elementary transitions;
moreover,  $A^{'}$ and $B^{'}$ (as well as A and B) can be groups of particles
with fixed (not growing with $s$) invariant masses. 

In the leading order the Reggeon vertices can be easily obtained from
the Feynman diagrams for the process $A + B \rightarrow A^{'} +
B^{'}$ in the Born approximation. Evaluating the diagrams we use
the light-cone momenta $p_1$ and $p_2$, so that in the general case
of massive particles  or clusters of particles $A$ and $B$ their 
momenta $p_A$ and $p_B$ are presented as
\begin{equation}
p_A = p_1 +\frac{m_A^2}{s}p_2 ~,~~~~ p_B = p_2
+\frac{m_B^2}{s}p_1~, ~~~~ s=2p_1p_2~.  
\label{z4}
\end{equation}
We use the Feynman gauge for virtual gluons; for external
gluons we use physical polarizations  with different gauge-fixing
conditions for gluons moving along $p_A$ and $p_B$, so that if
the gluon momentum  $k$  has large component along $p_1$ ($p_2$),
its polarization vector $e(k)$ satisfies  equations  $e(k)k=
e(k)p_2=0$ ( $e(k)k= e(k)p_1=0$). For the gluon propagator
connecting the vertices $\mu$ and $\nu$ with momenta predominantly
along $p_1$ and $p_2$ respectively  we do the usual trick of
retaining only the first term in the decomposition of the metric
tensor 
\begin{equation}
g^{\mu\nu} = \frac{2p_2^\mu p_1^\nu}{s}+\frac{2p_1^\mu
p_2^\nu}{s}+g_\perp^{\mu\nu} \rightarrow \frac{2p_2^\mu
p_1^\nu}{s}~
\label{z5}
\end{equation}
in the numerator of the propagator.  Using this trick one obtains
that in the leading order the Reggeon vertex $\Gamma _{A^{\prime
}A}^{(0)i}$ is equal to the $Ag\rightarrow A^{\prime }$ amplitude,
where the gluon $g$ has colour index $i$ and polarization
vector equal to $-p_2/s$. But in the next orders of perturbation
theory this relation is evidently broken. Moreover, contrary to
usual QCD vertices (such as, for example, the quark-quark-gluon
vertex) for which we  can draw a definite set of Feynman diagrams
with perfectly defined rules for the calculation of their
contributions, we have not such rules for the Reggeon vertices.
These vertices are extracted from the comparison of radiative corrections
to the $A + B \rightarrow A^{'} + B^{'}$ scattering amplitudes
with the Reggeized form expressed in Eq.~(\ref{z1}).  At the 
NLO the  Feynman diagrams for the process $A + B \rightarrow
A^{\prime} + B^{\prime}$ can be divided into four classes.  The first 
class includes  corrections to the $t$-channel gluon
propagator, the second and third are related to corrections to
the interaction of the $t$-channel gluon with the particles $A,
A^\prime$ and $B, B^\prime$ correspondingly, and the last one
contains the diagrams with the two-gluon exchange in the $t$-channel. 
The contributions of the diagrams of the first three classes have the same 
dependence on $s$ as the Born amplitudes; moreover, the contributions 
of the first and second (first and third) classes depend on 
properties of the particles $B, B^\prime$ $(A, A^\prime)$ in the same 
way as the Born amplitudes. It is evident therefore that the contribution 
of the diagrams of the second (third) class must be attributed to the vertex
$\Gamma _{A^{\prime }A}^{i}~(\Gamma _{B^{\prime }B}^{i})$,
whereas the contribution of the first class must be divided in
equal parts between these vertices. Consequently,  only the two-gluon 
exchange diagrams create a problem. In their contributions corrections 
to both vertices and the trajectory are mixed, so that the problem is 
to separate them. Note that there is a well known uncertainty in the 
vertices evident from Eq.~(\ref{z1}): we can change the vertices changing  simultaneously 
the energy scale.  The scale which we have chosen is $-t$, as it is  fixed in  
Eq.~(\ref{z1}).

Let us analyze the contributions of the two gluon exchange 
diagrams. They are shown schematically in Fig.1. Using  the Sudakov
decomposition for the gluon momenta:
\begin{equation}
k = \beta p_1 +\alpha p_2 +k_{\perp}~,~~~~~~ q = p_{A\prime} - p_A = 
p_B - p_{B\prime} =  \beta_q p_1 + \alpha_q p_2 +q_{\perp}~, 
\label{z6}
\end{equation}
we get for the Sudakov variables
\begin{equation}
\alpha_q = \frac{\vec q^{~2}+m_{A^\prime}^2-m_A^2}{s}~,
~~~~\beta_q = -\frac{\vec q^{~2}+m_{B^\prime}^2-m_B^2}{s}~,
~~~~d^{D}k = \frac{s}{2}d\alpha d\beta d^{D-2}k_{\perp}~.
\label{z7}
\end{equation}
Evaluating the diagrams of Fig.1 we retain, as usually,  only the
first term in the decomposition of the metric tensor (\ref{z5}).

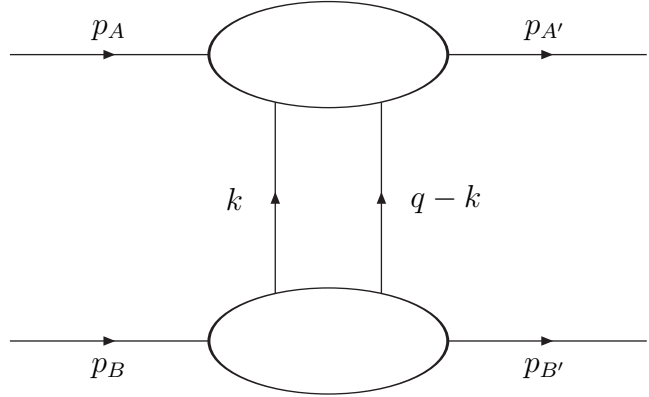
\begin{figure}[tb]      
\begin{center}
\begin{picture}(240,200)(0,0)

\ArrowLine(0,190)(75,190)
\ArrowLine(165,190)(240,190)
\Text(37.5,200)[]{$p_A$}
\Text(202.5,200)[]{$p_{A'}$}
\Oval(120,190)(20,45)(0)

\ArrowLine(100,100)(100,172)
\ArrowLine(140,100)(140,172)

\Text(85,136)[]{$k$}
\Text(165,136)[]{$q-k$}

\ArrowLine(0,82)(75,82)
\ArrowLine(165,82)(240,82)
\Text(37.5,72)[]{$p_B$}
\Text(202.5,72)[]{$p_{B'}$}
\Oval(120,82)(20,45)(0)

\end{picture}
\end{center}
\vspace{-2.0cm}
\caption[]{Schematic representation of diagrams with two-particle state 
in the $t$-channel for the process $A + B \rightarrow A' + B'$ at NLO.} 
\end{figure}

The first important observation is that in the region $|\alpha|
\ll 1 $ $~~(|\beta| \ll 1)$ we can factor out the vertex $\Gamma
_{B^{\prime }B}^{(0)i}$  $~~(\Gamma _{A^{\prime }A}^{(0)i})$ from
the diagrams of Fig.1. This is evident for the colour structure, since
we consider the colour octet and negative signature in the
$t$-channel, so that the virtual gluons in the $t$-channel must
be taken in the antisymmetric colour octet. The operator  $\hat
{{\cal P}}_{{8^-}}$ for the projection of the two-gluon colour
states  on the antisymmetric colour octet is given by
\begin{equation}
\langle c_1c_1^{\prime }|\hat {{\cal P}}_{8^-}|c_2c_2^{\prime
}\rangle =\frac{ f_{c_1c_1^{\prime }c}f_{c_2c_2^{\prime }c}}{N}~, 
\label{z8}
\end{equation}
where $f_{abc}$ are the structure constants of the colour group.
The relation
\begin{equation}
\frac{f_{c_2c_2^\prime
c}}{N}\left(T^{c_2^\prime}T^{c_2}\right)_{B^\prime B}=-
\frac{i}{2}T^{c}_{B^\prime B}~ 
\label{z9}
\end{equation}
defines the relevant colour coefficients. For the ordinary spin
structure this observation is almost evident as well in the case when the
particles $A$ and $A^\prime$ ($B$ and $B^\prime$) are quarks,
since the numerators of the quark propagators at the lower (upper)
parts of the diagram Fig.1 are surrounded by $\not{p}_1$
$(\not{p}_2)$ due to Eq.~(\ref{z5}). In the gluon case it is not
difficult to see too, though it is less evident. Let us consider,
for definiteness, the lower part of the diagram Fig.1, assuming
that the particles $B$ and $B^\prime$ are gluons. Remind that we
use the Feynman gauge for virtual gluons and physical
polarizations $e_B$ and $e_{B\prime}$ ($e_Bp_B=0=e_{B^\prime}
p_{B^\prime})$ in the gauge $e_Bp_1 = e_{B^\prime} p_1 = 0$.
Taking the polarization vectors of both $t$-channel gluons equal to 
$p_1/s$ according to Eq.~(\ref{z5}), we obtain the following expression 
for the lower part of the diagram of Fig.1 projected by the operator 
defined in Eq.~(\ref{z8}) on the antisymmetric colour octet state:
\[
{\cal
M}_{BB^\prime}^{c_1c_1^{\prime}}=\frac{g^2}{2}f_{c_1c_1^\prime
c}T^{c}_{B^\prime B}e_B^\beta \frac{p_1^\mu}{s}\frac{p_1^\nu}{s}
e_{B^\prime}^{\beta^\prime}\left[
\frac{{\gamma_{\beta\mu}}^{\rho}(p_B, -k, k-p_B)
\gamma_{\rho\nu\beta^\prime}(p_B-k,k-q,-p_{B^\prime})}{(p_B-k)^2}
\right.
\]
\begin{equation}
\left. - \frac{{\gamma_{\beta\nu}}^{\rho}(p_B, k-q, -k-p_{B^\prime})
\gamma_{\rho\mu\beta^\prime}(p_{B^\prime}+k,-k,-p_{B^\prime})}{(p_{B^\prime}+k)^2}
\right]~. 
\label{z10}
\end{equation}
Here $c_1$ and $c_1^\prime$ are the colour indices of the gluons
with momenta $k$ and $q-k$ correspondingly and
\begin{equation}
\gamma^{\mu\nu\rho}(k_1,k_2,k_3)=\left[g^{\mu\nu}(k_1-k_2)^\rho
+g^{\mu\rho}(k_3-k_1)^\nu+g^{\nu\rho}(k_2-k_3)^\mu \right]~
\label{z11}
\end{equation}
is the three-gluon vertex. The convolution in Eq.~(\ref{z10}) gives us
\[
{\cal
M}_{BB^\prime}^{c_1c_1^{\prime}}=\frac{g^2}{2}f_{c_1c_1^\prime
c}T^{c}_{B^\prime
B}\left(\frac{\left[-e_B^\rho(1-{\alpha}/{2})-2(e_Bk){p_1^\rho}/{s}\right]
\left[-e_{{B^\prime}\rho}(1-{\alpha}/{2})-2(e_{B^\prime}(k-q))
{p_{1\rho}}/{s}\right]}{(p_B-k)^2}\right.
\]
\[
\left.
-\frac{\left[-e_B^\rho(1+{\alpha}/{2})-2(e_B(q-k)){p_1^\rho}/{s}\right]
\left[-e_{{B^\prime}\rho}(1+{\alpha}/{2})+2(e_{B^\prime}(k))
{p_{1\rho}}/{s}\right]}{(p_{B^\prime}+k)^2}\right)=
\]
\begin{equation}
\frac{g^2}{2}f_{c_1c_1^\prime c}T^{c}_{B^\prime
B}(e_Be_{B^\prime})\left(\frac{{(1-{\alpha}/{2})}^2}{(p_B-k)^2}-
\frac{{(1+{\alpha}/{2})}^2}{(p_{B\prime}+k)^2}\right)~.
\label{z12}
\end{equation}
Since  $(e_Be_{B^\prime})=-\delta_{\lambda_B\lambda_{B^\prime}}$
and in the case under consideration the particles $B$ and
$B^\prime$ are massless gluons, so that $m_B=m_{B^\prime}=0$, 
the result at small $\alpha$ can be written as 
\begin{equation}
{\cal
M}_{BB^\prime}^{c_1c_1^{\prime}} = -\frac{g}{2}f_{c_1c_1^\prime
c}\left(\frac{1}{(p_B-k)^2-m_B^2}-\frac{1}{(p_{B\prime}+k)^2-m_{B^\prime}^2}\right)\Gamma
_{B^{\prime }B}^{(0)c} ~.    
\label{z13}
\end{equation}
It is easy to see that in  this form the obtained equation is
valid also for the case when the particles $B$ and $B^\prime$ are
quarks. It  completes the proof of the factorization of the vertex
$\Gamma _{B^{\prime }B}^{(0)i}$ at $|\alpha|\ll 1$ .

Let us consider now the basic integrals $I$ and $I^\prime$ for the
diagrams of Fig.1. The first of them is 
\begin{equation}
I=\int \frac{d^{D}k}{(2\pi
)^{D}i}\frac{1}{(k^{2}+i0)((q-k)^{2}+i0)
((p_{A}+k)^{2}-m_A^2+i0)((p_{B}-k)^{2}-m_B^2+i0)}~,  
\label{z14}
\end{equation}
and the second can be obtained from $I$ by the substitution $p_B\leftrightarrow
-p_{B^\prime}$. Following Ref.~\cite{FM98}, we introduce three
regions of the $\alpha$ and $\beta$ variables:
\begin{displaymath}
\mbox{the central region:}~~~~~~~~~~~~|\alpha| < \alpha_0~,~~~~~~
|\beta| < \beta_0~,
\end{displaymath}
\begin{displaymath}
\mbox{the A-region:}~~~~~~~~~~~~|\beta| \geq
\beta_0~,~~~~~~|\alpha| < \alpha_0~,
\end{displaymath}
\begin{equation}
\mbox{the B-region:}~~~~~~~~~~~~|\alpha| \geq \alpha_0~,~~~~~~
|\beta| < \beta_0~, 
\label{z15}
\end{equation}
$\alpha_0$ and $\beta_0$ being chosen so that
\begin{equation}
\alpha_0 << 1~,~~~~~~\beta_0 >> 1~,~~~~~~s\alpha_0\beta_0 >> |t|~.
\label{z16}
\end{equation}
As well as in Ref.~\cite{FM98} we will take the limit
$s\alpha_0\beta_0 \rightarrow \infty$, while $\alpha_0
\rightarrow 0$ and $\beta_0 \rightarrow 0$, when $s \rightarrow
\infty$. Our definition of the regions differs from that used in
Ref.~\cite{FM98}, but this difference concerns only the regions where
$|\alpha|$ and $|\beta|$ are both ``large'' ($|\beta| > \beta_0,
~~~(|\alpha|
>\alpha_0)$) and is therefore nonessential, since
these regions  give a contribution of relative order
$|t|/(s\alpha_0\beta_0)$ which vanishes  in the limit $s
\rightarrow \infty$ \cite{FM98}.  Consequently, it is sufficient to
calculate only the contributions of the three regions defined by the 
relations (\ref{z15}) and (\ref{z16}). 
Moreover, calculating the contribution of the region $A$ $(B)$ we
can remove the restriction on $\alpha$ $(\beta)$ that can simplify 
calculations.  It is clear that the contribution of the region $A$ 
$(B)$ after extracting the factor $2s\Gamma_{B^{\prime
}B}^{(0)i}/t ~~~(2s\Gamma _{A^{\prime }A}^{(0)i}/t)$ must be
attributed to the vertex $\Gamma _{A^{\prime }A}^{i}$
($\Gamma_{B^{\prime }B}^{i}$). Since in the $A$-region
\begin{equation}
(p_B-k)^2-m_B^2\simeq -(p_{B\prime}+k)^2+m_{B\prime}^2 \simeq
-2p_2k =-s\beta~, 
\label{z17}
\end{equation}
we obtain for the contribution of this region,  using Eqs.~(\ref{z8}),
(\ref{z9}) and (\ref{z13}) and restoring all relevant
coefficients,
\begin{equation}
\Gamma_{A^\prime A}^{i (\mbox{\scriptsize {A}})}=
g\frac{t}{2s}T^i_{c_1c_1^\prime}\int \frac{d^{D}k}{(2\pi
)^{D}i}\frac{p_2^\mu p_2^\nu
A_{\mu\nu}^{c_1c_1^\prime}(p_A,k;p_{A^\prime},k-q)}{(k^{2}+i0)((q-k)^{2}+i0)(p_2
k)}\theta(2|p_2k|-\beta_0s)~.  
\label{z18}
\end{equation}
Here $T^i_{ab}=-if_{iab}$ are the matrix elements of the colour
group generators in the adjoint representation,
$A_{\mu\nu}^{c_1c_1^\prime}(p_A,k;p_{A^\prime},k-q)$ is the one-particle irreducible 
in the $t$-channel part of the  amplitude of the
process $A+g(k)\rightarrow A^\prime +g(k-q)$. Moreover we used the
possibility, which we discussed above, to  remove the restriction $|\alpha|
\ll \alpha_0$ in the definition (\ref{z15}) of the region $A$.

In the central region the integrand in Eq.~(\ref{z14}) can be
considerably simplified.  First of all, evidently we can neglect
the longitudinal components of the momentum transfer $q$. Remind
that  the Reggeon vertices are calculated taking the limit $s
\rightarrow \infty$ before the limit $\epsilon \rightarrow 0$, so
that performing the analysis we can think of the transverse momenta
of the $t$-channel gluons being fixed (not depending on $s$).
Since $\beta_0 \rightarrow 0$ when $s \rightarrow \infty$, it is
easy to see that,  whereas the pole in the complex plane of
$\alpha$ from the third denominator in Eq.~(\ref{z14}) is found at
fixed values of $s|\alpha|$, all other poles are in the region
$s|\alpha| \rightarrow \infty$. Therefore the  contour of the
integration over $\alpha$ can be shifted to the region  $s|\alpha|
\rightarrow \infty$. An analogous conclusion is valid for the
integration over $\beta$. It means that we  can use the relations 
(\ref{z17}) in the central region too; moreover, along with these 
relations we can put also
\begin{equation}
(p_A+k)^2-m_A^2\simeq -(p_{A\prime}-k)^2+m_{A\prime}^2 \simeq
2p_1k =s\alpha~, 
\label{z19}
\end{equation}
so that  we obtain
\[
I^{\mbox{\scriptsize {central}}}=
\]
\[
\frac{s}{2}\int\frac{d^{D-2}k_\perp}{(2\pi
)^{D}i}\int_{-\alpha_0}^{\alpha_0}
\int_{-\beta_0}^{\beta_0}\frac{d\alpha d\beta}{(s\alpha \beta +
k_\perp^{2}+i0)(s\alpha\beta +(q-k)_\perp^{2}+i0) (s\alpha
+i0)(-s\beta+i0)}\simeq
\]
\begin{equation}
\frac{1}{2s}\int\frac{d^{D-2}k_\perp}{(2\pi
)^{D-1}(k_\perp^{2}-(q-k)_\perp^{2})}\left[
\frac{1}{k_\perp^{2}}\ln
\left(\frac{-s\alpha_0\beta_0}{-k_\perp^{2}}\right)-\frac{1}{(q-k)_\perp^{2}}\ln
\left(\frac{-s\alpha_0\beta_0}{-(q-k)_\perp^{2}}\right)\right]~.
\label{z20}
\end{equation}
The integrals over $k_\perp$ can be easily calculated and we get 
\cite{FM98}
\[I^{\mbox{\scriptsize
{central}}} = \frac{\Gamma (1-\epsilon )} {(4\pi )^{2+\epsilon
}}\frac{\Gamma ^{2}(\epsilon )}{\Gamma (2\epsilon )} \frac{(\vec
q^{~{2}})^{\epsilon }}{st} \left[ \ln \left(\frac{-s\alpha
_{0}\beta _{0}}{\vec q^{~{2}}}\right)-\psi (1)+\psi(1-\epsilon
)-2\psi (\epsilon )+2\psi (2\epsilon )\right]=
\]
\begin{equation}
-\frac{\omega^{(1)}(t)}{g^2Nst}\left[ \ln \left(\frac{-s}{\vec
q^{~{2}}}\right)+\phi(\alpha_0)+\phi(\beta_0)\right]~, 
\label{z21}
\end{equation}
where
\begin{equation}
\phi(z)=\ln z +\frac12\left(\frac1\epsilon
-\psi(1)+\psi(1-\epsilon)-2\psi(1+\epsilon)+2\psi(1+2\epsilon)\right).
\label{z22}
\end{equation}
Let us emphasize that this contribution does not depend on
masses of the particles $A$ and $B$. In the central region both
vertices $\Gamma _{A^{\prime }A}^{(0)i}$ and $ \Gamma _{B^{\prime
}B}^{(0)i}$ are factored out, so that we have
\[
A_{8^-}^{\mbox{\scriptsize {central}}}=-g^2Ns^2 \Gamma _{A^{\prime
}A}^{(0)i}\left[I^{\mbox{\scriptsize {central}}}-I^{\prime
~\mbox{\scriptsize {central}}}\right] \Gamma _{B^{\prime}B}^{(0)i}=
\]
\begin{equation}
\Gamma _{A^{\prime }A}^{(0)i}\frac{2s}{t}\omega^{(1)}(t)\left[
\frac12\ln \left(\frac{-s}{\vec q^{~{2}}}\right)+\frac12\ln
\left(\frac{s}{\vec
q^{~{2}}}\right)+\phi(\alpha_0)+\phi(\beta_0)\right] \Gamma
_{B^{\prime }B}^{(0)i}~.
 \label{z23}
\end{equation}
The above relation was proved in Ref.~\cite{FM98} for the case of the
process $\gamma^*q\rightarrow (q\bar q)q$, where the $q\bar q$
pair is produced in the fragmentation region of the photon. But,
since we have the gluon Reggeization (which was proved
\cite{BLF} in the LLA), and since the $s$-dependence can come only
from the two-gluon exchange diagrams, it is clear that this
relation is valid for any process. It can be proved for any
particular process applying the trick (\ref{z5}) to propagators
of both $t$-channel gluons. Clearly, the logarithmic terms in 
Eq.~(\ref{z23}) correspond to the expansion of the
power terms in Eq.~(\ref{z1}), whereas all other contributions must
be distributed between the corrections to the vertices $\Gamma
_{A^{\prime }A}^{i}$ and $ \Gamma _{B^{\prime }B}^{i}$. The way
to do it is evident,  so that
\begin{equation}
\Gamma_{A^\prime A}^{i (\mbox{\scriptsize {central}})}=\Gamma
_{A^{\prime }A}^{(0)i} \omega^{(1)}(t)\phi(\beta_0)~.
 \label{z24}
\end{equation}
The intermediate parameter $\beta_0$ in Eq.~(\ref{z24}) cancels when we
combine this contribution with the contribution to  $\Gamma _{A^{\prime
}A}^{i}$ from the A-region (\ref{z18}).

Let us summarize what we have obtained. To the one loop accuracy
the Reggeon vertex $\Gamma _{A^{\prime }A}^{(i)}$ can be presented
as a sum of the following contributions:

  -the  $Ag\rightarrow A^{\prime }$ amplitude, where the virtual 
gluon $g$  has the colour index $i$
and the polarization vector $-p_2/s$, with the gluon self-energy
taken with the coefficient $1/2$, just as for external particles;

   -the sum of the contributions given by Eqs.~(\ref{z18}) and (\ref{z24}). 
Remind that $A_{\mu\nu}^{c_1c_1^\prime}(p_A,k;p_{A^\prime},k-q)$ in Eq.~(\ref{z18}) 
is the one-particle irreducible in the $t$-channel part of the
amplitude of the process $A+g(k)\rightarrow A^\prime +g(k-q)$ in
the leading order.

The essential point is that in this approach one can use known
results for the gluon propagator and vertices, and the only new 
piece which must be calculated is 
$\Gamma_{A^\prime A}^{i(\mbox{\scriptsize {A}})}$, given by 
Eq.~(\ref{z18}), which can be easily found (see below).

\subsection{Reggeized quark vertices}
The same approach can be used for the calculation of effective
vertices of the Reggeon-particle interaction in the case when the
Reggeon is the Reggeized quark. Due to the quark Reggeization the 
amplitudes with the quark quantum numbers in the $t$-channel  and 
the positive signature can be presented~\cite{FS} in a way analogous 
to the form (\ref{z1}):
\begin{equation}
A_{3^+}=\Gamma _{A^{\prime }A}\frac{1}{m-\not
\!q_{\perp}}\frac{1}{2}\left[ \left( \frac{-s}{-t}\right)
^{\delta(\not q_{\perp})}+\left( \frac{s}{-t}\right)
^{\delta(\not q_{\perp})}\right] \Gamma _{B^{\prime }B}~,
\label{z25}
\end{equation}
where $m$ is the quark mass, $\Gamma _{A^{\prime }A}$ are the vertices 
for the $A\rightarrow A^{\prime}$ transitions and 
$\delta(\not \!\!q_{\perp})$ determines the quark Regge trajectory. 
In the leading order~\cite{FS}
\[
\delta(\not \!q_{\perp})=\delta^{(1)}(\not
\!q_{\perp})=\frac{g^{2}}{(2\pi )^{D-1}}C_F(\not \!q_{\perp}-m)\int
\frac{d^{D-2}k_{\perp}}{({\not k_{\perp}-m})(q-k)_{\perp}^{2}}=
\]
\begin{equation}
\frac{g^2\Gamma(1-\epsilon)}{(4\pi)^{2+\epsilon}}2C_F(\not{\!q}_\perp-m)
\int_0^1\frac{dx(x\not{\!q}_\perp+m)}{((1-x)(m^2+\vec
q^{~2}))^{1-\epsilon}} ~, 
\label{z26}
\end{equation}
where $C_F=(N^2-1)/(2N)$. Elementary transitions due the to coupling
with the Reggeized quark in the leading order are:

- the gluon $\rightarrow$ quark transition with the vertex
\begin{equation}
\Gamma ^{(0)}_{QG}=-g\bar u(p_Q)\not{\!e}(p_G)t^G ~,  
\label{z27}
\end{equation}
where $p_Q$ and $p_G$ are the quark and gluon momenta, $e(p_G)$ is
the gluon polarization vector and $t^G$ is the colour group generator in
the fundamental representation corresponding to the gluon colour state,

- the gluon $\rightarrow$ antiquark transition with the vertex
\begin{equation}
\Gamma^{(0)}_{\bar Q G}=-g\not{\!e}(p_G)t^G v(p_{\bar Q})~,  
\label{z28}
\end{equation}

- the corresponding vertices for inverse transitions which are
defined by
\begin{equation}
\Gamma^{(0)}_{GQ}=\bar \Gamma^{(0)}_{QG} \equiv
\gamma^0\left(\Gamma^{(0)}_{QG}\right)^\dag~,\,\,\,\Gamma_{G\bar
Q}=\bar \Gamma^{(0)}_{\bar Q G}\equiv \left(\Gamma^{(0)}_{\bar Q 
G}\right)^\dag \gamma^0~. 
\label{z29}
\end{equation}
Remind that for  gluons moving  along $p_1$ $(p_2)$ their
polarization vectors  must be taken in the gauge $e(p_G)p_2=0$
$(e(p_G)p_1=0)$.

The above vertices can be  easily obtained  from the Feynman
diagrams for the process $A + B \rightarrow A^{'} + B^{'}$ with
the quark exchange in the $t$-channel in the Born approximation. 
Note, for definiteness, that $q = p_{A\prime} - p_A = 
p_B - p_{B\prime}$ is the momentum flowing along the quark line, and 
that the form (\ref{z25}) does not take into account the factor (-1) 
prescribed by the Feynman rules in the case when both the beginning 
and the end of the quark line correspond to the antiquark. So, in the 
leading order the  vertices for the Reggeized quark
simply coincide with the vertices for the ordinary quark, with  
properly chosen gluon polarizations.  Of course, it is not the
case in  the next orders of perturbation theory.  At the NLO,
quite analogously to the case of the Reggeized gluon, the 
corresponding Feynman diagrams are divided into four classes. The
first class corresponds to corrections to the $t$-channel quark
propagator, the second and third are related to corrections to the
interaction of the $t$-channel quark with the particles $A,
A^\prime$ and $B, B^\prime$ respectively, and the last one
contains the diagrams with the quark-gluon exchange in the
$t$-channel. The contribution of the diagrams of the second
(third) class must be attributed to the vertex $\Gamma _{A^{\prime
}A}^{i}$ $ (\Gamma _{B^{\prime }B}^{i})$, whereas the contribution
of the first class must be divided in equal parts between these
vertices. Again  the only uncertainty comes from the two-particle
exchange diagrams, schematically presented in Fig.1.  

Again we use the Sudakov decomposition (\ref{z6}), retain only the 
first term in the decomposition of the metric tensor in the $t$-channel 
gluon propagator (\ref{z5}) and consider the 
three regions (\ref{z15}) of the $\alpha$ and $\beta$ variables. In the
region $|\alpha| < \alpha_0$ $~~(|\beta| < \beta_0)$ we can factor
out the vertex $\Gamma _{B^{\prime }B}^{(0)}$ $~~(\Gamma
_{A^{\prime }A}^{(0)})$ from the diagrams of Fig.1. The relevant
coefficients in the colour space can be calculated using the
operator $\hat {{\cal P}}_{{3}}$ for the projection of the
quark-gluon colour states in the $t-$channel on the fundamental
representation:
\begin{equation}
\langle c\alpha|\hat {{\cal P}}_{3}|c^\prime\alpha^\prime\rangle
=\langle\alpha|\frac{t^ct^{c^\prime}}{C_F}|\alpha^\prime\rangle~ 
\label{z30}
\end{equation}
and the relations
\begin{equation}
t^ct^bt^c=\left (C_F-\frac{N}{2}\right )t^b~,~~~~~~~~~
t^ct^iT^c_{ib}=\frac{N}{2}t^b~. 
\label{z31}
\end{equation}

Unlike the case of the two-gluon exchange, where use  of
the operator  $\hat {{\cal P}}_{{8^-}}$ automatically gives us
the negative signature amplitude, here we need to perform
"signaturization" in an explicit way.

Let us consider the lower part of the diagram of Fig.1 in the case when the
particle $B$ is the quark and $B^\prime$ is the gluon. Without
loss of generality we can take the exchanged  particle with
momentum $k$ for the quark. Then the particle with momentum $q-k$ is
the gluon; let its colour index be $c$. Denoting 
${\cal M}_{B^\prime B}^{c\beta^\prime}$ the matrix element of the 
lower part of the diagram of Fig.1, with $p_1/s$ instead of its 
polarization vector (see Eq.~(\ref{z5})) and having omitted the  
wave functions $u(p_B)$ and $e_{B^\prime}^{\beta^\prime}$ 
of the external particles, we obtain
\[
{\cal M}_{B^\prime B}^{c\beta^\prime} = ig^2\left[\frac{p_{1\nu}}{s}
\gamma^{\rho\nu\beta^\prime}(p_{B}-k,k-q,-p_{B^\prime})\gamma_\rho
\frac{[t^{B^\prime},t^c]}{(p_B-k)^2}-\gamma^{\beta^\prime}\frac{(\not
p_{B^\prime}+\not k+m)}{(p_{B^\prime}+k)^2-m^2}\frac{\not
p_1}{s}t^{B^\prime}t^c\right]\simeq
\]
\begin{equation}
-ig^2\gamma^{\beta^\prime}\left[\frac{[t^{B^\prime},t^c]}{(p_B-k)^2}
(1-\frac\alpha2)+\frac{t^{B^\prime}t^c}{(p_{B^\prime}+k)^2-m^2}
(1+\alpha)\right]~.
\label{z32}
\end{equation}
Here $\gamma^{\rho\nu\beta^\prime}$ is the three-gluon vertex 
defined in Eq.~(\ref{z11}). 
We have taken into account that ${\cal M}_{B^\prime B}^{c\beta^\prime}$ 
must be convoluted with the polarization vector
$e_{B^\prime}^{\beta^\prime}$ of the gluon $B^\prime$ satisfying
$e_{B^\prime}p_1=0$, so that we have omitted the terms with
$p_1^{\beta^\prime}$. We have omitted also the terms with ${\not
p_1}/{s}$ standing on the most left position. This can be done
because in such a case we can move, with the help of the
anticommutation relations for $\gamma$ matrices, ${\not p_1}/{s}$
to act on the spinor from the upper blob of Fig.1 and then use
the Dirac equation, getting only terms vanishing at large $s$.

Since we include in Eq.~(\ref{z32}) an extra dependence on $s$ through 
the polarization vector of the $t$-channel gluon, in order to take 
the amplitude with the positive signature we must antisymmetrize the 
last expression with respect to the substitution
$p_B\leftrightarrow -p_{B^\prime}$. In the $A$-region
($|\alpha|<\alpha_0, ~~\beta \geq \beta_0$) it gives us
\begin{equation}
e_{B^\prime}^{\beta^\prime}{\cal M}_{B^\prime
B}^{(+)c\beta^\prime}u(p_B) =
-i\frac{g^2}{2}t^c\left[\frac{1}{2p_2k}-\frac{1}{-2p_2k} \right]
t^{B^\prime}\not\!{e}_{B^\prime}u(p_B)~, 
\label{z33}
\end{equation}
that proves  the factorization of the vertex 
$\Gamma^{(0)}_{B^\prime B}=-gt^{B^\prime}{\!{\not{\!e}_{B^\prime}}}u(p_B)$
in the case when the particle $B$ is the quark and $B^\prime$ is the
gluon. Pay attention that the matrix $t^c$ in Eq.~(\ref{z33}) performs 
the projection of the quark-gluon state in the $t$-channel on the fundamental
representation (see Eq.~(\ref{z30})), so that only this representation 
does survive in the positive signature at 
small $|\alpha|$. The  case  when the particle $B$ is the gluon and 
$B^\prime$ is the antiquark can be considered quite similarly. As a 
result one obtains the  factorization of the vertex with
$\Gamma^{(0)}_{B^\prime B}$ with the same coefficient as in 
Eq.~(\ref{z33}).

Note that although the two terms in the square brackets in 
Eq.~(\ref{z33}) are equal in the $A$-region, they are written
separately. It is done in order to have a possibility to use this
expression in the central region as well. The basic integrals for
the considered case of the  quark-gluon state in the $t$-channel
differ in the central region from the corresponding integrals for the
two-gluon state because of the non zero quark mass and the numerator 
$\not{k}_\perp+m$ of the quark propagator; however it does not change 
the properties of the integrals which were discussed after 
Eq.~(\ref{z18}). The contour of integration over $\beta$ in the central 
region can be shifted to the region $s|\beta| \rightarrow \infty$, so 
that the approximations used in the derivation of Eq.~(\ref{z33}) can 
be used in the central region too;  the only difference is that in the
central region we have to take into account the positions of the poles 
in the complex $\beta$-plane, which are fixed by the $+i0$ prescription 
in the denominators of Eq.~(\ref{z33}) and are different for the 
two terms there.

As well as in the Reggeized gluon case, the contribution of the region 
$A$ must be attributed to the vertex $\Gamma _{A^{\prime }A}$. Extracting 
the factor $(m-{\!{\not{\!q}_\perp)^{-1}}}\Gamma_{B^{\prime }B}^{(0)}$
from the total contribution of this region, we obtain 
\begin{equation}
\Gamma_{A^\prime A}^{(\mbox{\scriptsize {A}})}=-g \int
\frac{d^{D}k}{(2\pi )^{D}i}\frac{p_2^\nu
A_{\nu}^{c}(p_A,k;p_{A^\prime},k-q)t^c}{(k^{2}-m^2+i0)((q-k)^{2}+i0)(p_2
k)}\theta(2|p_2k|-\beta_0s)(\not{k}+m)(m-\not{\!q}_\perp)~.
\label{z34}
\end{equation}
Here  $A_{\nu}^{c}(p_A,k;p_{A^\prime},k-q)$ is the one-particle 
irreducible in the $t$-channel part of the amplitude of the
process $A+q(k)\rightarrow A^\prime +g(k-q)$, $c$ is the colour
index of the gluon with momenta $k-q$ and $\nu$ is its
polarization index. As well as in Eq.~(\ref{z18}) we removed here the
restriction $|\alpha| \ll \alpha_0$ in the definition of the
region $A$ (see Eq.~(\ref{z15}).

The contribution of the central  region can be presented as
\begin{equation}
A_{3^+}^{\mbox{\scriptsize {central}}}=g^2C_Fs\Gamma _{A^{\prime
}A}^{(0)}\left[J^{\mbox{\scriptsize {central}}}-J^{\prime
~\mbox{\scriptsize {central}}}\right] \Gamma _{B^{\prime
}B}^{(0)}~,
\label{z35}
\end{equation}
where, as it was already mentioned, the integrals
$J^{\mbox{\scriptsize {central}}}$ and $J^{\prime~
{\mbox\scriptsize {central}}}$ differ only  by the quark mass 
and the numerator $\not{k}_\perp+m$ in the integrands from the 
integrals $I^{\mbox{\scriptsize {central}}}$  (see Eq.~(\ref{z20})) 
and $I^{\prime ~\mbox{\scriptsize {central}}}$, so that
\[
J^{\mbox{\scriptsize {central}}}
=
\]
\[
\frac{s}{2}\int\frac{d^{D-2}k_\perp}{(2\pi
)^{D}i}\int_{-\alpha_0}^{\alpha_0}
\int_{-\beta_0}^{\beta_0}\frac{d\alpha d\beta
(\not{k}_\perp+m)}{(s\alpha \beta +
k_\perp^{2}-m^2+i0)(s\alpha\beta +(q-k)_\perp^{2}+i0) (s\alpha
+i0)(-s\beta+i0)}
\]
\begin{equation}
\simeq
\frac{1}{2s}\int\frac{d^{D-2}k_\perp(\not{k}_\perp+m)}{(2\pi
)^{D-1}(k_\perp^{2}-m^2-(q-k)_\perp^{2})}\left[
\frac{1}{k_\perp^{2}-m^2}\ln
\left(\frac{-s\alpha_0\beta_0}{m^2-k_\perp^{2}}\right)-\frac{1}{(q-k)_\perp^{2}}\ln
\left(\frac{-s\alpha_0\beta_0}{-(q-k)_\perp^{2}}\right)\right]~.
\label{z36}
\end{equation}
Of course, this contribution does  not depend on masses of the 
particles $A$ and $B$. The integral $J^{\prime~ \mbox{\scriptsize
{central}}}$ differs from $J^{\mbox{\scriptsize {central}}}$ only
by the sign of $s$.

Integration over $k_\perp$ yields
\begin{equation}
J^{\mbox{\scriptsize {central}}} =
\frac{\delta^{(1)}(\not{\!q}_\perp)}{g^2C_F2s(m-\not{\!q}_\perp)}\ln\left(\frac{-s}{\vec
q^{~2}}\right)+\frac{1}{2s}(\Delta(\not{\!q}_\perp, \alpha_0)
+\Delta(\not{\!q}_\perp, \beta_0))~,
\label{z37}
\end{equation}
where $\delta(\not{\!q}_\perp)$ is defined in Eq.~(\ref{z26}) and 
\[
\Delta(\not{\!q}_\perp,z)=
\]
\begin{equation}
\frac{\Gamma(1-\epsilon)}{(4\pi)^{2+\epsilon}}
\int_0^1\frac{dx(x\not{\!q}_\perp+m)}{((1-x)(m^2+x\vec
q^{~2}))^{1-\epsilon}}\left(\psi(1)-\psi(1-\epsilon)+\ln\left(\frac{(1-x)(m^2+x\vec
q^{~2})}{\vec q^{~2}z^2}\right)\right) ~.  
\label{z38}
\end{equation}
Consequently, for the contribution of the central region (\ref{z35}) we
find 
\[
A_{3^+}^{\mbox{\scriptsize {central}}}=
\]
\begin{equation}
\Gamma _{A^{\prime}A}^{(0)}\left[\frac{\delta(\not{\!q}_\perp)}
{2(m-\not{\!q}_\perp)}\left(\ln\left(\frac{-s}{\vec
q^{~2}}\right) +\ln\left(\frac{s}{\vec
q^{~2}}\right)\right)+g^2C_F (\Delta(\not{\!q}_\perp, \alpha_0)
+\Delta(\not{\!q}_\perp, \beta_0)\right] \Gamma _{B^{\prime
}B}^{(0)}~.
\label{z39}
\end{equation}
This relation is valid for any process. The terms proportional to
$\delta(\not{\!\!q}_\perp)$  in Eq.~(\ref{z39}) correspond to the expansion
of the Regge factors  in Eq.~(\ref{z25}); the remaining terms
 must be distributed between the corrections to the
vertices $\Gamma _{A^{\prime }A}$ and $ \Gamma _{B^{\prime
}B}$. Therefore we obtain
\begin{equation}
\Gamma_{A^\prime A}^{(\mbox{\scriptsize {central}})}=\Gamma
_{A^{\prime }A}^{(0)}g^2C_F
(m-\!\not{\!q}_\perp)\Delta(\not{\!q}_\perp, \beta_0)~,
 \label{z40}
\end{equation}
where $\Delta(\not{\!q}_\perp, \beta_0)$ is defined in Eq.~(\ref{z38}).
The parameter $\beta_0$ in the contribution~(\ref{z40}) cancels when we
add to it the contribution (\ref{z34})  from the A-region.

As a result, we obtain the prescription for the calculation of
the Reggeon vertices to the one loop accuracy. The vertex $\Gamma
_{A^{\prime }A}$ turns out to be the sum of the following
contributions:

  -the  $Aq\rightarrow A^{\prime }$ amplitude, with the quark self-energy taken
with the coefficient $1/2$, just as for external particles;

   -the contributions given by Eqs.~(\ref{z34}) and \ref{z40}). Remind 
that $A_{\nu}^{c}(p_A,k;p_{A^\prime},k-q)$ is the one-particle irreducible 
in $t$-channel part of the amplitude of the process $A+q(k)\rightarrow 
A^\prime +g(k-q)$, $c$ is the colour index of the gluon with momentum 
$k-q$ and $\nu$ is its polarization index. The parameter $\beta_0$
cancels in the sum of these contributions.

It is important that in this approach the bulk of the vertex is expressed
in terms of the quark self energy and vertices which are known. The only 
piece which must be calculated is 
$\Gamma_{A^\prime A}^{(\mbox{\scriptsize {A}})}$, defined by Eq.~(\ref{z40}).

Note that for massless quarks we get 
\begin{equation}
\delta^{(1)}(\not{\!q}_\perp)=-\frac{g^2C_F\Gamma(1-\epsilon)}{(4\pi)^{2+\epsilon}}\left(\vec
q^{~2}\right)^\epsilon\frac{\Gamma^2(\epsilon)}{\Gamma(2\epsilon)}=\frac{C_F}{N}\omega^{(1)}(t)~,
\label{z41}
\end{equation}
\[
\Delta(\not{\!q}_\perp,
z)=\frac{\Gamma(1-\epsilon)}{(4\pi)^{2+\epsilon}}\not{\!q}_\perp\left(\vec
q^{~2}\right)^{\epsilon-1}\frac{\Gamma^2(\epsilon)}{2\Gamma(2\epsilon)}\left(-2\ln
z+ \psi(1)-\psi(1-\epsilon)
+2\psi(\epsilon)-2\psi(2\epsilon)\right)
\]
\begin{equation}
=\frac{\not{\!q}_\perp}{\vec
q^{~2}}\frac{\omega^{(1)}(t)}{g^2N}\phi(z)~, 
\label{z42}
\end{equation}
where $\omega^{(1)}(t)$ is defined in Eq.~(\ref{z2}) and $\phi(z)$ in
Eq.~(\ref{z22}). For the corrections to the vertex from the central
region in the massless case we obtain from Eq.~(\ref{z40})
\begin{equation}
\Gamma_{A^\prime A}^{(\mbox{\scriptsize {central}})}=\Gamma
_{A^{\prime }A}^{(0)}\frac{C_F\omega^{(1)}(t)}{N}\phi(\beta_0)~,
 \label{z43}
\end{equation}
with $\phi(z)$ defined in Eq.~(\ref{z22}).

\section{Calculation of the  Vertices}
Let us demonstrate how the method can be applied calculating the 
one-loop corrections to the vertices of interaction 
of the Reggeized gluon and quark with particles $A$ and
$A^\prime$, which are ordinary quarks and gluons. In the
following we consider massless quarks, so that the momenta of the
particles $A$ and $A^\prime$ are equal to $p_1$ and $p_{1^\prime}$ 
respectively and the Reggeon momentum is
\begin{equation}
q=p_{1^\prime}-p_1=-\frac{\vec q^{~2}}{s}p_1+\frac{\vec
q^{~2}}{s}p_2+q_\perp\simeq q_\perp~. 
\label{z44}
\end{equation}
Considering the Reggeized gluon we denote its colour index by $c$.

\subsection{Quark-Quark-Reggeized gluon  Vertex}
According to our prescription, the first contribution comes from the
amplitude $q(p_1)+g(q)\rightarrow q(p_{1^\prime}$), with the gluon
polarization vector equal to $-p_2/s$. The one-loop
corrections to this amplitude are well known. Putting the 
Reggeon vertex in the form
\begin{equation}
\Gamma^c_{Q^\prime Q}=\Gamma^{(0)c}_{Q^\prime
Q}(1+\delta_Q)~,
\label{z45}
\end{equation}
where the Born vertex $\Gamma^{(0)c}_{Q^\prime Q}$ is defined in
Eq.~(\ref{z3}), and taking into account that
\begin{equation}
\bar
u(p_{1^\prime})\frac{\not{p}_2}{s}u(p_1)=
\delta_{\lambda_{1^\prime}\lambda_1}~,
\label{z46}
\end{equation}
we obtain from the gluon self-energy (taken with the coefficient
$1/2$)
\begin{equation}
\delta^{s.e.}_Q=\omega^{(1)}(t)\frac{(5+3\epsilon)N
-2(1+\epsilon)n_f}{4(1+2\epsilon)(3+2\epsilon)N}
\label{z47}
\end{equation}
and from the irreducible quark-quark-gluon vertex
\begin{equation}
\delta^{v}_Q=\omega^{(1)}(t)\left[\frac{-1}{4(1+2\epsilon)}
-\frac{1}{4N^2}\left(1+\frac{2}{\epsilon(1+2\epsilon)}\right)\right]~.
\label{z48}
\end{equation}
Here $n_f$ is the number of quark flavors and $\omega^{(1)}(t)$ is
defined in Eq.~(\ref{z2}). The contribution of the quark self-energy is
zero for massless quarks.

The only contribution which we need to calculate is defined by 
Eq.~(\ref{z18}):
\[
\Gamma^{c(A)}_{Q^\prime Q}=
\]
\begin{equation}
g^3N\frac{t}{s}u(p_{1^\prime})\int\frac{d^{D}k}{(2\pi)^Di}
\frac{\not{p}_2(\not{p}_1+\!\not{k})\not{p}_2t^c}{(k^{2}+i0)((q-k)^{2}+i0)
((p_1+k)^2+i0)(2p_2k)}\theta(2|p_2k|-\beta_0s)u(p_1)~.
\label{z49}
\end{equation}
Using the Sudakov decomposition (\ref{z6}) and neglecting $\beta_q$ 
with respect to $\beta$ we obtain
\[
\delta^{A}_Q=g^2N\frac{t}{2}\int \frac{d\beta}{\beta}
(1+\beta)\theta(|\beta|-\beta_0)\int\frac{d^{D-2}k_\perp}{(2\pi)^{D-1}}
\]
\begin{equation}
\times \int\frac{d(s\alpha)}{2\pi i}\frac{1}{(s\alpha \beta +
k_\perp^{2}+i0)(s(\alpha-\alpha_q)\beta +(q-k)_\perp^{2}+i0)
(s\alpha(1+\beta)+k_\perp^2 +i0)}. 
\label{z50}
\end{equation}
The integral over $\alpha$ can be easily calculated with the residue
method. Then, changing $\beta$ with $-\beta$  we have
\[
\delta^{A}_Q=g^2N\frac{t}{2}\int_{\beta_0}^1 \frac{d\beta}{\beta}
(1-\beta)^2\int\frac{d^{D-2}k_\perp}{(2\pi)^{D-1}}\frac{1}{k_\perp^{2}
(k_\perp-q_\perp(1-\beta))^2}=
\]
\begin{equation}
-g^2N\frac{\Gamma(1-\epsilon)}{(4\pi)^{2+\epsilon}}\left(\vec
q^{~2}
\right)^\epsilon\frac{\Gamma^2(\epsilon)}{\Gamma(2\epsilon)}\int_{\beta_0}^1
\frac{d\beta}{\beta} (1-\beta)^{2\epsilon}
=\omega^{(1)}(t)\left(-\ln \beta_0
+\psi(1)-\psi(1+2\epsilon)\right)~. 
\label{z51}
\end{equation}
The total correction is the sum of the pieces  given by 
Eqs.~(\ref{z47}), (\ref{z48}), (\ref{z51}) and  the contribution 
(\ref{z24}) from the central region, with $\phi(z)$ 
defined in Eq.~(\ref{z22}). The result we arrive at is
\[
\delta_Q=\omega^{(1)}(t)\frac12\biggl[\frac{1}{\epsilon} +
\psi(1-\epsilon) + \psi(1) - 2\psi(1+\epsilon)+
\frac{2+\epsilon}{2(1+2\epsilon)(3+2\epsilon)}
\]
\begin{equation}
 -\frac{1}{2N^2}\left(1 + \frac{2}
{\epsilon(1+2\epsilon)} \right) -
\frac{n_f}{N}\frac{(1+\epsilon)}{(1+2\epsilon)(3+2\epsilon)}
\biggr]~,  
\label{z52}
\end{equation}
in accordance with Ref.~\cite{FFQ}.

\subsection{Gluon-Gluon-Reggeized gluon Vertex}
Let us denote with $e_1$ and $e_{1^\prime}$ the polarization vectors 
of the gluons with momenta $p_A=p_1$ and $p_{A^\prime}=p_{1^\prime}$ 
correspondingly; they satisfy the conditions $e_1p_1=
e_{1^\prime}p_{1^\prime}=0~,~~ e_1p_2= e_{1^\prime}p_2=0$.

Since the helicity conservation in the Born vertices (see 
Eq.~(\ref{z4})) is violated by radiative corrections \cite{FL93}, let
us write the Reggeon vertex in a general form:
\begin{equation}
\Gamma^c_{G^\prime G}=gT^c_{G^\prime
G}\left[\delta_{\lambda_{1^\prime}\lambda_1}(1+\delta^{(+)}_G)
+\delta_{\lambda_{1^\prime},-\lambda_1}\delta^{(-)}_G\right]~,
\label{z53}
\end{equation}
where $T^c_{G^\prime G}$ are the matrix elements of the colour
group generator in the adjoint representation and 
\begin{equation}
\delta_{\lambda_{1^\prime}\lambda_1}=-e_{1^\prime}^*e_1~,
~~~~~~~\delta_{\lambda_{1^\prime},-\lambda_1}=-e_{1^\prime}^*e_1
+(D-2)\frac{(e_{1^\prime}^*q)(e_1q)}{q^2}~.
\label{z54}
\end{equation}

The first contribution to the vertex comes from the amplitude
$g(p_1)+g(q)\rightarrow g(p_{1^\prime})$, where the polarization
vector of the gluon with  momentum $q=p_{1^\prime}-p_1$ is taken
equal to $-p_2/s$. The one-loop corrections to this amplitude are
well known. From the gluon self-energy, taken with the
coefficient $1/2$, we have a contribution only to $\delta^{(+)}_G$,
the same as to $\delta_Q$ of Eq.~(\ref{z47}):
\begin{equation}
\delta^{(+)(s.e.)}_G)=\omega^{(1)}(t)\frac{(5+3\epsilon)N
-2(1+\epsilon)n_f}{4(1+2\epsilon)(3+2\epsilon)N}~. 
\label{z55}
\end{equation}
From the irreducible three-gluon vertex we have contributions to
both helicity conserving and non-conserving parts; moreover, for
the helicity non-conserving part there are no other
contributions, so that we have
\begin{equation}
\delta^{(+)v}_G=\omega^{(1)}(t)\left[\frac{3}{8\epsilon}+\frac{1}{2(1+\epsilon)}
-\frac{5}{4(1+2\epsilon)}-\frac{1}{3+2\epsilon}
+\frac{n_f}{2N}\frac{2(1+\epsilon)^3+\epsilon^2}
{(1+\epsilon)^2(1+2\epsilon)(3+2\epsilon)}\right]~,
\label{z56}
\end{equation}
\begin{equation}
\delta^{(-)}_G=\frac{\epsilon\omega^{(1)}(t)}{2(1+\epsilon)(1+2\epsilon)(3+2\epsilon)}
\left(-1+\frac{n_f}{N(1+\epsilon)}\right)~, 
\label{z57}
\end{equation}
Remind that  $n_f$ is the number of quark flavors and 
$\omega^{(1)}(t)$ is defined in Eq.~(\ref{z2}). Contributions from
the self-energy of the on-mass shell gluons  are zero.

The only contribution which we need to calculate is defined by 
Eq.~(\ref{z18}). We get for it:
\[
\Gamma^{c(A)}_{G^\prime G}=- g^3NT^c_{G^\prime G}\frac{t}{2s}\int
\frac{d^{D}k}{(2\pi )^{D}i} \frac{p_2^\mu p_2^\nu
}{(k^{2}+i0)((q-k)^{2}+i0)(p_2 k)}\theta(2|p_2k|-\beta_0s)
\]
\begin{equation}
\times \frac{e_{1^{\prime}}^{* \alpha^\prime}e_{1}^{\alpha}
{\gamma_{\alpha\mu}}^{\rho}(p_1,k,-p_1-k)\gamma_{\rho\nu\alpha^\prime}
(p_1+k,k-q,-p_{1^\prime})}{(p_1+k)^2}~, 
\label{z58}
\end{equation}
where the vertices  $\gamma_{\mu\nu\rho}$ are defined in 
Eq.~(\ref{z11}). The calculation of this contribution can be done in the
same way as in the $\Gamma^{c(A)}_{Q\prime Q}$ case. Using the
Sudakov decomposition (\ref{z6}) and neglecting $\beta_q$ with 
respect to $\beta$ we obtain
\[
\Gamma^{c(A)}_{G^\prime G}=- g^3NT^c_{G^\prime G}e_1^*e_1\frac{t}{2}\int
\frac{d\beta}{\beta} (1+\beta /2
)^2\theta(|\beta|-\beta_0)\int\frac{d^{D-2}k_\perp}{(2\pi)^{D-1}}
\]
\begin{equation}
\times \int\frac{d(s\alpha)}{2\pi i}\frac{1}{(s\alpha \beta +
k_\perp^{2}+i0)(s(\alpha-\alpha_q)\beta +(q-k)_\perp^{2}+i0)
(s\alpha(1+\beta)+k_\perp^2 +i0)},  
\label{z59}
\end{equation}
so that we find
\[
\delta^{(+)A}=g^2N\frac{t}{2}\int_{\beta_0}^1
\frac{d\beta}{\beta}(1-\beta /2)^2
(1-\beta)\int\frac{d^{D-2}k_\perp}{(2\pi)^{D-1}}\frac{1}
{k_\perp^{2}(k_\perp-q_\perp(1-\beta))^2}
\]
\[
=-g^2N\frac{\Gamma(1-\epsilon)}{(4\pi)^{2+\epsilon}}\left(\vec
q^{~2}
\right)^\epsilon\frac{\Gamma^2(\epsilon)}{\Gamma(2\epsilon)}\int_{\beta_0}^1
\frac{d\beta}{\beta} (1-\beta)^{2\epsilon-1}(1-\beta /2)^2
\]
\begin{equation}
=\omega^{(1)}(t)\left(-\ln \beta_0
+\psi(1)-\psi(1+2\epsilon)+\frac{1}{8\epsilon(1+2\epsilon)}\right)~.
\label{z60}
\end{equation}
The total correction to the helicity conserving vertex is given by
the sum of the contributions (\ref{z55}), (\ref{z56}), (\ref{z60}) and 
(\ref{z24}), the last from the central region, with  
$\phi(z)$ defined in Eq.~(\ref{z22}). We obtain
\[
\delta^{(+)}_G=\omega^{(1)}(t)\frac12\biggl[\frac{2}{\epsilon} +
\psi(1-\epsilon) + \psi(1) - 2\psi(1+\epsilon)-
\frac{9(1+\epsilon)^2+2}{2(1+\epsilon)(1+2\epsilon)(3+2\epsilon)}
\]
\begin{equation}
+
\frac{n_f}{N}\frac{(1+\epsilon)^3+\epsilon^2}{(1+\epsilon)^2(1+2\epsilon)(3+2\epsilon)}
\biggr]~,  
\label{z61}
\end{equation}
in accordance with Ref.~\cite{FL93}.

\subsection{Gluon-Quark-Reggeized quark Vertex}
Calculation of one-loop corrections to the vertices of the
Reggeized quark is performed in a similar way. Let us 
consider for definiteness the vertex for the gluon $\rightarrow$
quark transition. The gluon and quark momenta are  $p_A=p_1$ and
$p_{A^\prime}=p_{1^\prime}$ correspondingly, the Reggeized quark
momentum is $q=p_{1^\prime}-p_1$ and  the gluon colour index is $G$. 
The gluon polarization vector $e$ satisfy the conditions $ep_1=
ep_2=0$. Then the vertex can be presented as
\begin{equation}
\Gamma_{QG}=-g\bar{u}(p_{1^\prime})t^G\left[\not{\!e}(1+\delta_e)
+\frac{(eq)\not{\!q}_\perp}{{q}^2_\perp}\delta_q\right]~.
\label{z62}
\end{equation}
The first contribution to the vertex comes from the
amplitude $g(p_1)+q(q)\rightarrow q(p_{1^\prime})$ where the quark 
with momentum $q$ is off-mass shell . Calculating this amplitude we 
can omit the terms with ${\not p_2}/{s}$ standing on the most right 
position, since in such case  we can move ${\not p_2}/{s}$ in the
Reggeized amplitude (\ref{z25})  with the help of the
anticommutation relations for $\gamma$ matrices to act on the
spinor wave function of the quark with momentum along $p_2$ and then
use the Dirac equation, getting only terms vanishing at large $s$.

The off-mass shell quark self-energy contributes only to
$\delta_e$. Taking into account  the coefficient $1/2$, we obtain
\begin{equation}
\delta_e^{s.e.}=\delta^{(1)}(\not{\!q}_\perp)\frac{1}{4}
\left(\frac{-1-\epsilon}{1+2\epsilon}\right)~.
\label{z63}
\end{equation}
Notice that for the massless quark
$\delta^{(1)}(\not{\!\!q}_\perp)=(C_F/N)\omega^{(1)}(t)$, being 
$\omega^{(1)}(t)$ defined in Eq.~(\ref{z2}). The contributions of the
self-energy of the on-mass shell quark and gluon are zero. The
irreducible gluon-quark-quark vertex contributes to both spin
structures:
\begin{equation}
\delta^{v}_e=\omega^{(1)}(t)\left[\frac{1}{2\epsilon}
-\frac{1}{4N^2}\left(1-\frac{2}{1+2\epsilon}\right)\right]~,
\label{z64}
\end{equation}
\begin{equation}
\delta^{v}_q=\omega^{(1)}(t)\left[\frac{-1}{2\epsilon}
+\frac{\epsilon}{2(1+2\epsilon)}-\frac{1}{2N^2}\left(\frac{2-\epsilon}{1+2\epsilon}\right)\right]~.
\label{z65}
\end{equation}

The unknown  contribution is defined by Eq.~(\ref{z34}):
\[
\Gamma_{QG}^{(\mbox{\scriptsize
{A}})}=g^3\bar{u}(p_{1^\prime})t^G\int \frac{d^{D}k}{(2\pi
)^{D}i}\frac{\theta(2|p_2k|-\beta_0s)}{(k^{2}+i0)((q-k)^{2}+i0)(p_2
k)}\left[\frac{1}{2N}\frac{\not{p}_2(\not{p}_1+\not{k})\not{\!e}}
{(p_1+k)^2+i0}\right.
\]
\begin{equation}
\left. +\frac{N}{2}\frac{p_2^\nu
e^\alpha\gamma^\rho\gamma_{\alpha\rho\nu}(p_1, -p_{1^\prime}+k,
q-k )}{(p_{1^\prime}-k)^2+i0}\right] \not{k}\not{\!q}_\perp~,
\label{z66}
\end{equation}
where the vertex $\gamma_{\mu\nu\rho}$ is defined in Eq.~(\ref{z11}).

The calculation of this contribution differs from the calculation
of $\Gamma_{Q^\prime Q}^{c(\mbox{\scriptsize {A}})}$ and
$\Gamma_{G^\prime G}^{c(\mbox{\scriptsize {A}})}$ only by the 
spinor and tensor algebra. The result is

\begin{equation}
\delta^A_e =\omega^{(1)}(t)\left[\frac{c_F}{N}\left(-\ln \beta_0
+\psi(1)-\psi(1+2\epsilon)\right)-\frac{1}{4(1+2\epsilon)}\right]~,
\label{z67}
\end{equation}
\begin{equation}
\delta^A_q
=\omega^{(1)}(t)\left[\frac{1}{2\epsilon}+\frac{1}{N^2(1+2\epsilon)}\right]~.
\label{z68}
\end{equation}
The contribution to the vertex from the central region is shown in 
Eq.~(\ref{z43}). The total correction $\delta_e$, given by the sum of 
the contributions (\ref{z63}), (\ref{z64}), (\ref{z67}) and the 
correction (\ref{z43})  from the central region, is
\begin{equation}
\delta_e
=\omega^{(1)}(t)\left[\frac{c_F}{2N}\left(\frac{1}{\epsilon}-
\frac{3(1-\epsilon)}{2(1+2\epsilon)}+\psi(1)+\psi(1-\epsilon)-
2\psi(1+\epsilon)\right)+\frac{1}{2\epsilon}-\frac{\epsilon}{2(1+2\epsilon)}\right]~,
\label{z69}
\end{equation}
\begin{equation}
\delta_q
=\omega^{(1)}(t)\frac{\epsilon}{2(1+2\epsilon)}\left(1+
\frac{1}{N^2}\right)~.
\label{z70}
\end{equation}
Note that $\delta_q$ is not singular at $\epsilon \rightarrow 0$, since the
corresponding spin structure in Eq.~(\ref{z62}) is absent at the leading order.
\footnotetext{We were informed by the authors of Ref. \cite{KLPV}, who
are considering this  vertex for the massive quark case in the $t-$channel 
unitarity approach \cite{FL93}, that in the massless limit their result  
is in complete agreement with our one. }

\section{Discussion}
We developed the method which permits both for the case of the Reggeized
gluon and the Reggeized quark to calculate the Reggeon-particle vertices at
the next-to-leading order directly, without calculation of scattering
amplitudes. In this order the only uncertainty in the definition of the
vertices is related to the Feynman diagrams with two particles in the
$t$-channel. The remarkable property of the integrals, corresponding to
these diagrams, is that they can be presented as the sum of the
contributions of three integration regions: the central region,
where the $t$-channel particles have small components of momenta along
the momentum of any of the colliding particles A and B, and two other
regions, which can be called regions of fragmentation of the particles A
and B respectively. It is possible to separate precisely these regions and
to define unambiguously their contributions. The contributions of the
fragmentation regions must be attributed to the corresponding vertices. As
for the ``central'' region, it occurs that its contribution does not depend
on the colliding particles. We want to stress that this
property is crucial for the Reggeization, so that we could conclude about its
existence from the fact of the Reggeization. It permits to determine the
contributions to the Reggeon-particle vertices from the central region,
which are universal (i.e. independent of the colliding particles), and to
formulate the rules of calculation of the vertices. For the vertices of the
Reggeized gluon these rules are given at the end of Subsection 2.1, for 
the vertices of the Reggeized quark at the end of Subsection 2.2.

Using these rules we calculated at the next-to-leading order the 
elementary vertices for the interaction of the Reggeized gluon and quark. 
The vertices of the Reggeized gluon are already known, so that in this 
case the calculations we performed serve for the demonstration of the 
validity of the method and the check of the results. As far as 
we know, the vertices of the Reggeized quark were not yet  
calculated in the NLO, so that these results are new. 

Note once more that the developed method is valid not only for the 
calculation of  the elementary transition vertices considered here. 
Evidently, it can be applied in the general case, including transitions 
between groups of particles with fixed (not growing with $s$) invariant masses. 

\vskip 1.5cm \underline {Acknowledgment}: V.S.F thanks the 
Dipartimento di Fisica della Universit\`a della Calabria and the Istituto
Nazionale di Fisica Nucleare - Gruppo collegato di Cosenza for their warm
hospitality while part of this work was done.

\end{document}